%% file: Ferdman_0537_arxiv.tex
\documentclass[twocolumn]{aastex61}
\pdfoutput=1 
\usepackage{amsmath,amstext,amssymb}
\usepackage[T1]{fontenc}
\usepackage{hyperref}
\usepackage{todonotes}
\usepackage[figure,figure*]{hypcap}
\usepackage{natbib}
\usepackage{latexsym}
\usepackage{ifthen}
\usepackage{url}
\usepackage{comment}

\newcommand{\rxte}{\textit{RXTE}}

\newcommand{\degrees}{\,^\circ}

\newcommand{\nudot}{\dot{\nu}}
\newcommand{\nuddot}{\ddot{\nu}}

\newcommand{\psr}[1]
{\ifthenelse{\equal{#1}{0540}}{PSR~B0540$-$69}{\ifthenelse{\equal{#1}{0531}}{PSR~B0531+21}{\ifthenelse{\equal{#1}{crab}}{PSR~B0531+21}{\ifthenelse{\equal{#1}{0537}}{PSR~J0537$-$6910}{\ifthenelse{\equal{#1}{1756}}{PSR~J1756$-$2251}{\ifthenelse{\equal{#1}{1906}}{PSR~J1906+0746}{\ifthenelse{\equal{#1}{1141}}{PSR~J1141$-$6545}{\ifthenelse{\equal{#1}{2127}}{PSR~B2127+11C}{\ifthenelse{\equal{#1}{0751}}{PSR~J0751+1807}{\ifthenelse{\equal{#1}{1713}}{PSR~J1713+0747}{ ???????}}}}}}}}}}}

\shorttitle{Timing behavior of PSR~J0537$-$6910}
\shortauthors{Ferdman et al.}

\begin{document}

\title{The glitches and rotational history of the highly energetic young pulsar PSR~J0537$-$6910}

\author{R.~D.~Ferdman} 
\affiliation{Faculty of Science, University of East Anglia, Norwich NR4 7TJ, United Kingdom; r.ferdman@uea.ac.uk}
\affiliation{Department of Physics \& McGill Space Institute, McGill University, 3600 University Street, Montreal, QC H3A 2T8, Canada}
\author{R.~F.~Archibald} 
\affiliation{Department of Physics \& McGill Space Institute, McGill University, 3600 University Street, Montreal, QC H3A 2T8, Canada}
\affiliation{Department of Astronomy and Astrophysics, University of Toronto, 50 St. George Street, Toronto, ON M5S 3H4, Canada}         
\author{K.~N.~Gourgouliatos} 
\affiliation{Department of Applied Mathematics, University of Leeds, Leeds LS2 9JT, United Kingdom}
\affiliation{Department of Mathematical Sciences, Durham University, Durham, DH1 3LE, United Kingdom}
\author{V.~M.~Kaspi} 
\affiliation{Department of Physics \& McGill Space Institute, McGill University, 3600 University Street, Montreal, QC H3A 2T8, Canada}


\begin{abstract}
We present a timing and glitch analysis of the young X-ray pulsar \psr{0537}, located within the Large Magellanic Cloud, using 13 years of data from the now decommissioned \textit{Rossi X-ray Timing Explorer}. Rotating with a spin period of 16\,ms, \psr{0537} is the fastest spinning and most energetic young pulsar known.  It also displays the highest glitch activity of any known pulsar.  We have found 42 glitches over the data span, corresponding to a glitch rate of 3.2\,yr$^{-1}$, with an overall glitch activity rate of $8.8\times 10^{-7}\,$yr$^{-1}$.  The high glitch frequency has allowed us to study the glitch behavior in ways that are  inaccessible in other pulsars.  We observe a strong linear correlation between spin frequency glitch magnitude and wait time to the following glitch. We also find that the post-glitch spin-down recovery is well described by a single two-component model fit to all glitches for which we have adequate input data.  This consists of an exponential amplitude $A = (7.6 \pm 1.0)\times 10^{-14}\,$s$^{-2}$ and decay timescale $\tau = 27_{-6}^{+7}\,$d, and linear slope $m = (4.1\pm 0.4)\times 10^{-16}\,$s$^{-2}\,$d$^{-1}$. The latter slope corresponds to a second frequency derivative $\ddot{\nu} = (4.7\pm 0.5) \times 10^{-22}\,$s$^{-3}$, from which we find an implied braking index $n=7.4 \pm 0.8$. We also present a maximum-likelihood technique for searching for periods in event-time data, which we used to both confirm previously published values and determine rotation frequencies in later observations. We discuss the implied constraints on glitch models from the observed behavior of this system, which we argue cannot be fully explained in the context of existing theories. 
\end{abstract}


\section{Introduction}
Observations of young pulsars provide a unique probe of the formation and early evolution of neutron stars (NSs), as well as insight into the physics governing their exotic, turbulent interiors and their magnetospheric activity. 
Precision timing analysis to  characterize long-term rotational evolution is one way to accomplish such studies.

Rotational irregularities, which are prevalent in young pulsars, are particularly telling.
Significant departures from a simple spin-down model are most often due to so-called ``timing noise'' and/or rotational glitch activity.  

Timing noise is the apparently random, long-term (i.e. $\sim$years) variations in the pulsar rotation frequency, thought to be caused by instabilities in the NS magnetosphere \citep[see, e.g.,][]{hlk10}.  Indeed, profile shape changes have been correlated with timing noise in several pulsars \citep{lhk+10}.

Glitches are near-instantaneous changes in spin frequency, often accompanied by a corresponding change in frequency derivative. They are thought to be caused by a transfer of angular momentum from an interior superfluid component to the stellar crust, which undergoes a rapid decoupling \citep[e.g.,][]{bppr69,ai75,hm15}.
Glitches are typically followed by a recovery period that can return the pulsar asymptotically to its pre-glitch spin parameters \citep{lsp92}, though it is sometimes found that the recovery overshoots this value \citep[e.g.][]{lkg10} or persists in its post-glitch frequency derivative value after a short recovery period \citep[e.g.,][]{ymh+13}.  
In-depth studies of pulsar glitch rates and magnitudes, as well as the recoveries that follow, have the potential to reveal the physical nature of the turbulent processes that take place in the interiors of NSs, and the matter that composes them.  

More specifically, glitch rate and the size of corresponding changes in $\nu$ and $\dot{\nu}$ are thought to be related to the age and interior temperature of the NS \citep[e.g.,][]{accp96, lel99}.  The oldest observed population of NSs, the millisecond pulsars (MSPs), are rarely seen to glitch, with only two MSP glitches seen to date \citep{cb04, mjs+16}.  On the other extreme, the very youngest pulsars ($\lesssim 2\,$kyr) tend to have lower glitch activity than slightly older NSs. For example, the Crab pulsar and \psr{0540} (characteristic age $\tau_c \equiv P/2\dot{P} \sim 1.3$ and $1.7\,$kyr, respectively) have been observed to have consistently small glitches \citep[][respectively]{ljg+15,fak15}, while the similarly aged PSR~B1509$-$58 ($\tau_c \sim 1.7\,$kyr) has exhibited no glitches in over 28 years of data \citep{lk11}. In contrast, the Vela pulsar, with $\tau_c \sim 11\,$kyr, is among the most actively glitching pulsars, with relative glitch sizes in frequency $\Delta\nu/\nu \gtrsim 5-10$ times larger than in the Crab, and more often by approximately three orders of magnitude \citep{elsk11}.  

Timing observations that can detect glitches are also sometimes able to measure an important property of a young pulsar: its braking index.  This can be measured via the second derivative of the pulsar frequency, $\ddot{\nu}$.  The dimensionless braking index is defined as $n=\nu \ddot{\nu}/\dot{\nu}^2$, a key characteristic of the pulsar spin-down evolution \citep{gm69}.  Different models of pulsar spin-down predict different values of $n$.
For example, $n=1$ corresponds to a purely particle wind-driven spin-down, $n=3$ to a dipole magnetic field being solely responsible for the rotational braking, and $n=5$ is a case where quadrupolar effects determine the rate at which the pulsar slows its rotation.  There are currently 11 measured braking indices \citep[see, e.g.,][and references therein]{llrg17}.  They vary between low values, as in the \citet{lpgc96} measurement of the Vela pulsar \citep[$n=1.4\pm 0.2$; however see][]{slk+16}, or as in the high magnetic-field pulsar PSR~J1734$-$3333 \citep[$n=0.9\pm 0.2$;][]{elk+11}, to relatively high values, as in PSR~J1640$-$4631 \citep[$n=3.15\pm 0.03$;][]{agf+16}. 

Braking indices are also seen to change, sometimes dramatically so: in \psr{0540}, it was observed that $\Delta n \sim -2.1$ after an apparently rapid phase change that resulted in a $36\%$ increase in the spin-down rate \citep{mgh+16}.  In PSR~J1846$-$0258, a change $\Delta n \sim -0.5$ was found to accompany a magnetar-like outburst \citep{lnk+11,akb+15}, and a $15\%$ decrease in braking index was found to accompany a glitch in PSR~J1119$-$6127 \citep{awe+15}.  These make clear that there remains much to understand about the physics of braking indices. 

\psr{0537} was discovered in the Large Magellanic Cloud, serendipitously as part of a search in SN~1987A with the \textit{Rossi X-ray Timing Explorer} (\textit{RXTE}) for a NS counterpart \citep{mgz+98}.  Associated instead with the nearby N157B supernova remnant in the 30~Doradus star-forming region, \psr{0537} is the most energetic known young pulsar:  it has the fastest rotation frequency of any young pulsar, $\nu = 62\,$Hz (corresponding to a rotation period $P=16\,$ms), and the highest spin-down luminosity, with $\dot{E} = 4.9\times 10^{38}\,$ergs\,s$^{-1}$.  It has a derived surface dipolar magnetic field of $B_s = 9.25\times 10^{11}\,$G, similar to other young NSs, and a characteristic age of 4.9\,kyr.

\psr{0537} immediately demonstrated evidence for glitch activity \citep{mgz+98}.  Indeed, long-term follow-up observations over 7.5 yr showed this to be the case:  by a large margin, it is the pulsar with the largest known glitch rate, $\sim 3\,$yr$^{-1}$, and the magnitudes of these glitches are among the largest observed among young pulsars \citep[][the latter henceforth referred to as M06]{mgm+04,mmw+06}. 

This substantial number of large glitches presents an opportunity to investigate NS interior physics in ways that are not possible in other pulsars.
For instance, evidence of a correlation has been observed between glitch size and the time until the following glitch (M06), leading to a crust-quake interpretation \citep{hm15}.  However, this relationship is not observed in other pulsars. 

The high glitch rate results in an overall decreasing trend in (but increase in the magnitude of) the pulsar spin-down rate over time, leading to a long-term \emph{negative} braking index $n=-1.5$ found by M06, and more recently $n=-1.2$ \citep{els16}.  It is unclear whether this gives a direct clue into the intrinsic behavior of the magnetosphere.  Instead, this may be an artifact of the glitch activity, which serves to inhibit the full spin-down recovery.  We discuss the effect of the high glitch rate of \psr{0537} on the overall spin evolution and the braking index in \S\ref{sec:braking_index}.  

Despite dedicated, deep searches, no significant radio emission has be observed from \psr{0537} \citep{cmj+05}.  In contrast, point-source emission coincident with \psr{0537} has been observed in $\gamma$-rays with the \textit{Fermi Large Area Telescope}, though to date no significant pulsed emission has been found \citep{fermi15}.  X-ray observations therefore provide the only current means of detailed studies of the spin evolution of this pulsar.

In this work, we present an analysis of \psr{0537} using the original M06 data set and remaining 5.5\,yr of the \textit{RXTE} observations, for a total of 13 yr of data.  In \S\ref{sec:obs}, we outline the observations and data reduction, including pulse arrival time determination, and the introduction of a novel method for finding pulse periods in event data such as those obtained by most X-ray telescopes.  In \S\ref{sec:timing} we describe our timing analysis, including the derivation of solutions corresponding to each inter-glitch wait time, as well as the long-term spin evolution of this pulsar throughout this data set, including a measurement of its nominal braking index.  In \S\ref{sec:glitches} we present our analysis of the glitches in \psr{0537}, which supports the existence of a strong correlation between glitch size and wait time.  We also derive an empirical two-component model for the glitch recoveries, which appears to follow a single functional description.  We also determine the evolution of the glitch activity parameter. \S\ref{sec:rad_change} briefly describes our search for radiative variations in this pulsar, over the long term, or as short-lived bursts or pulse profile changes coincident with the glitches. In \S\ref{sec:discussion} we discuss the physical interpretation of our findings, and their implications for our understanding of glitch and post-glitch behavior in this pulsar, its braking index, and how these relate to the overall picture of magnetospheric and interior evolution in this and other NSs.  Finally, in \S\ref{sec:conclusions} we offer concluding thoughts about this work and prospects for future study of this unique pulsar.

\section{Observations and data set}
\label{sec:obs}
We have extended the M06 data set to include a total span of 13 years of observations for this work. 
We analyzed data taken exclusively by the \textit{RXTE} Proportional Counter Array \citep[PCA;][]{jmr+06}.  
The five proportional counter units (PCUs) that comprised the PCA together give an approximately $1\degrees$ field of view over a 6500~cm$^2$ field of view. Typically, fewer than five PCUs were simultaneously operating, and so we used data from all active PCUs during a given observation. 
The PCA observations result in ``event'', data files; these are a collection of time stamp/energy pairs corresponding to individual candidate photons that have triggered the PCU detectors.

In order to find the optimal range of event energies for our data set, we constructed pulse profiles based on the known pulsar ephemeris.  For this, we used event data near in time to the reference epoch of the ephemeris, so as to avoid smearing or drifting effects due to, e.g. glitch activity, which would affect the calculated pulse phase of the events.
We found that of the range of energies to which the PCA is sensitive ($2-60\,$keV over 256 spectral channels), the pulse profile signal-to-noise ratio (S/N) was maximized in the range $3-20\,$keV; we therefore use this range for our timing analysis.
The data were taken in ``GoodXenon'' mode, which provided 1-$\mu$s time resolution.

In order to correct the event times to barycentric dynamical time (TDB), we used the \texttt{barycorr} script provided with the \textit{RXTE} {\tt FTOOLS} package, part of the {\tt HEASOFT} software suite\footnote{see \url{http://heasarc.gsfc.nasa.gov/lheasoft/ftools/fhelp/barycorr.html}}.  This was done using the most recently observed and most precise position for \psr{0537}, found by \citet{tbf+06}: $\alpha$= 05:37:47.416; $\delta$= -69:10:19.88 (J2000).  This position was also used and held fixed throughout our timing analysis described in later sections.

\begin{figure}
  \begin{center}
    \includegraphics[width=0.5\textwidth]{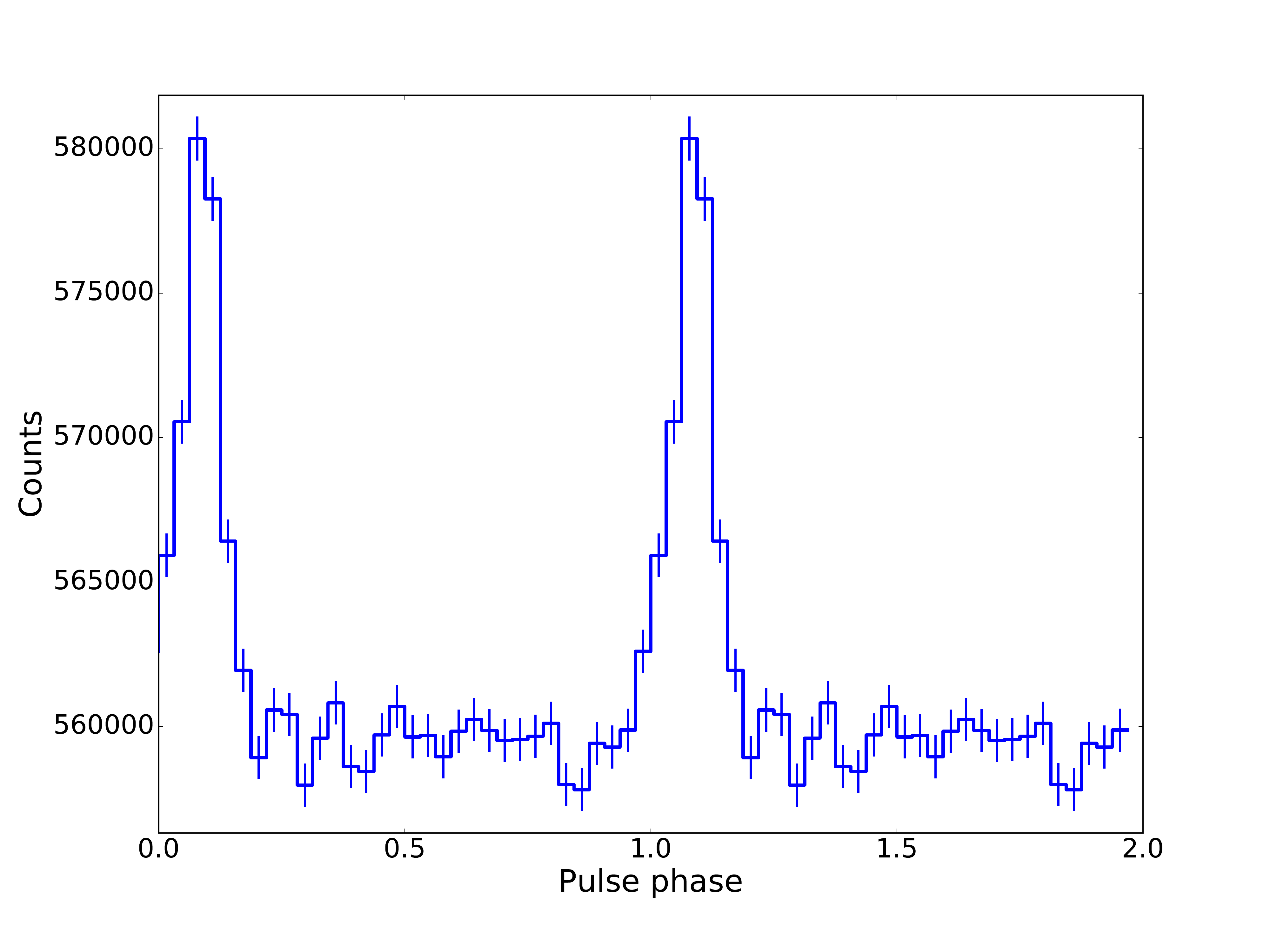}
    \caption{Template pulse profile in the 3--20 keV range used for maximum-likelihood TOA calculation. Two pulse periods are shown for clarity.\label{fig:template_prof}}
  \end{center}
\end{figure}

\subsection{Period search}
\label{sec:period_search}

Due to the frequent glitches observed in \psr{0537}, combined with the relatively sparse observing cadence and low source count rate, establishing phase-coherent solutions was challenging for the extended data set (i.e. after the final date reported in the M06 analysis). 
In order to determine the pulse period at the epoch of each \rxte{} observation, we developed a novel extension of a maximum-likelihood (ML) technique used to calculate pulse times of arrival (TOAs), which we briefly describe here and in \S\ref{sec:ml_toa}.  A full description is found in \citet{lrc+09}, who first introduced this method.  

We first created a high S/N template profile.  This was constructed from data taken between Modified Julian Dates (MJDs) 51986 and 52136, which forms one of the longest-duration and most sampled inter-glitch segments in our data set.  The template profile was accumulated by assigning each event to one of 32 rotational phase bins, according to an initial pulsar spin ephemeris taken from \citet{mgm+04}, and using the above-mentioned pulsar position from \citet{tbf+06}. The resulting profile is shown in Fig.~\ref{fig:template_prof}.

For this period search technique, we fit the high S/N template profile shown in Figure~\ref{fig:template_prof} with a single-component Gaussian, which we used as a likelihood function $I(\Phi)\ (0 \leq \Phi \leq 1)$.
A phase $\phi_i$ is calculated for each of $N$ events, based on the best model pulsar ephemeris. 
A trial phase offset $\Delta\phi$ is added to this set of phases, which is a function of both trial reference TOA $T$ and trial rotation period $P$.
The resulting two-dimensional probability density is then given by: 
\begin{equation}
	\label{eqn:2d_pdf}
	\mathrm{Prob}(T, P) = \prod_{i=1}^N I\left(\phi_i(P) - \delta\phi(T, P)\right)\,.
\end{equation}
We evaluate Equation~\ref{eqn:2d_pdf} over a finely sampled trial period grid.  For each period, we calculate the corresponding event phases and trial phase offsets. 
This results in a two-dimensional probability density function (PDF) over $P$ and $T$, from which we determine the median period and uncertainty for each event file by marginalizing over $T$.  
For this work, we took a typical grid range of $P_{\mathrm{trial}} \pm 3\times 10^{-4}$\,ms with $1\times 10^{-7}$\,ms resolution, centered on a linear fit to the ephemeris values from M06, with the period derivative held at the mean of these solutions \footnote{This grid spacing is based on the typical data set duration of 1 day, and on the 1\% phase measurements possible for phase alignment.}.  We determined the rotation period in this way over the 13 yr of data used for this work, including the reprocessing of the data set used by M06.  These periods were used as input for further timing analysis.

\subsection{Times of arrival}
\label{sec:ml_toa}
In contrast to other bright X-ray pulsars such as PSR~B0540$-$69, the other young pulsar in the Large Magellanic Cloud \citep[e.g.,][]{lkg05,fak15}\footnote{At angular distance of only $16\arcmin$ away, PSR~B0540$-$69 is present in the same \rxte{} data sets used in this work.}, individual integrated pulses from \psr{0537} have relatively low S/N. The traditional method of frequency-domain cross-correlation with our template profile \citep{tay92} gives often insufficiently precise or inaccurate TOAs for achieving phase-coherent solutions, particular during sparsely sampled segments of data. 
This would in turn have severely reduced the significance of our partially coherent timing parameters.  

To remedy this, we utilized the ML algorithm as described in \citet{lrc+09}, 
which directly uses the event data, and therefore does not lose information due to binning into evenly sampled time series, as is needed for Fourier or other periodogram-based techniques. 
As described in \S\ref{sec:period_search}, it utilizes the unbinned model of the pulse profile as a likelihood function. 
However, rather than determining the period as in \S\ref{sec:period_search}, a phase $\phi_i$ is calculated for each of $N$ events based on the periods found in our search. 
The trial phase offset $\Delta\phi$ is now only a function of trial TOA $T$, and the resulting probability density determined for our phase offsets is given by:
\begin{equation}
	\label{eqn:1d_pdf}
    \mathrm{Prob}(t_0) = \prod_{i=1}^N I(\phi_i - \delta\phi(T)).
\end{equation}
Calculating this over a sufficiently fine grid of trial phase offsets between 0 and 1 gives us a one-dimensional PDF in phase offset, and therefore TOA $T = P\Delta\phi$, where $P$ is the pulsar rotation period at time $T$.

\input{0537_par_table}

\section{Timing Analysis}
\label{sec:timing}
For all timing analyses, we use the \textsc{tempo2} software package \citep{hem06, ehm06}, which fits our barycentered TOAs to a model describing the evolution of the rotational phase of the pulsar.  This evolution can be described by a Taylor expansion of the spin frequency $\nu$ and frequency derivatives as a function of time $t$:
\begin{equation}
	\phi(t) = \phi(t_0) + \nu_0(t - t_0) + \frac{1}{2}\nudot_0(t - t_0)^2 + \frac{1}{6}\nuddot_0(t - t_0)^3 + \ldots,
\end{equation}
where $\nu_0$, $\nudot_0$, and $\nuddot_0$ are the reference frequency and its derivatives, respectively, at reference epoch $t_0$.
For our analysis, we held fixed the most recent known position for \psr{0537} reported by \citet{tbf+06}, and used the DE421 Solar System model from the Jet Propulsion Laboratory \citep{sta04b} in order to account for the motion of the Earth.  

With the TOAs found in as described in \S\ref{sec:ml_toa}, we derived individual phase-coherent solutions for each inter-glitch data segment, the results of which are shown in Table~\ref{tab:0537_subset_solutions}.  Glitch epochs were taken to be the midpoint in time between the final TOA of the preceding data segment and the first TOA of the following subset.
Uncertainties were taken to be the time difference between the glitch epoch and the nearest neighboring pairs of TOAs. We found a total of 42 glitches between MJDs 51197 and 55927, giving an average glitch rate of $3.24\,$yr$^{-1}$. A summary of these glitches is provided in Table~\ref{tab:0537_glitches}.  

\begin{figure*}
  \begin{center}
    \includegraphics[width=\textwidth]{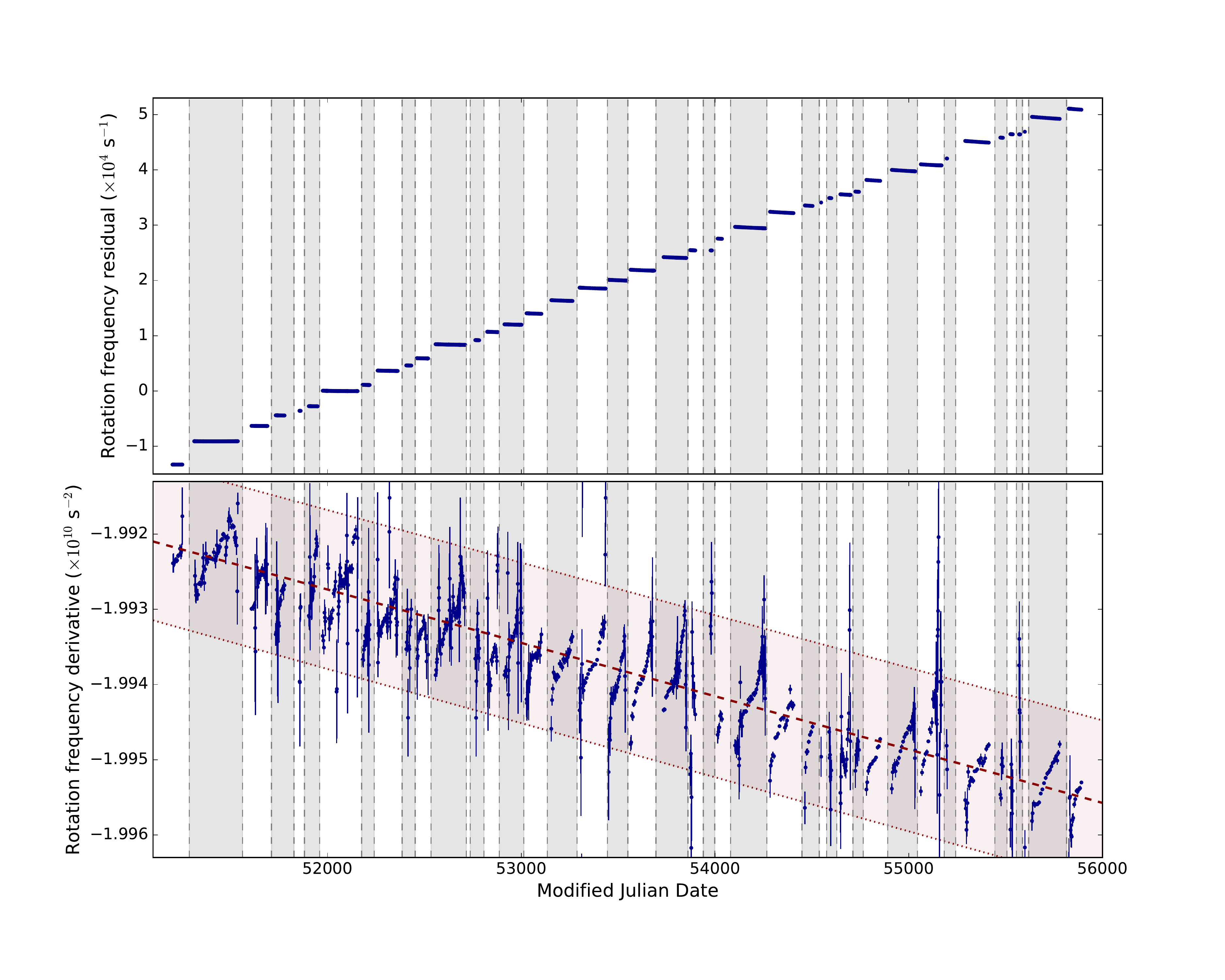}
    \caption{Partially coherent timing of PSR~J0537$-$6910, showing calculated rotation frequency residuals (\textit{top}) and frequency derivatives (\textit{bottom)} over time. Glitch epochs are denoted by grey dashed lines, with inter-glitch regions filled with alternating white and grey background. The red dashed line and faint red region represent the overall best-fit median second frequency derivative $\ddot{\nu}$ and 68\% confidence interval, corresponding to a formal braking index of $-1.28 \pm 0.04$ (but see text for details).\label{fig:partial_coherent}}
  \end{center}
\end{figure*}

\subsection{Short-Term Timing}
The unusually high glitch activity of \psr{0537}, together with its long-term quasi-random timing noise, renders our ability to perform traditional long-baseline coherent timing analysis difficult.
To mitigate these effects, we determine the short-term timing behavior between glitches by using the technique of \emph{partially coherent} timing \citep[e.g.][]{lkg05, fak15}.
To do so, we extracted the absolute TOA pulse numbers from each inter-glitch timing solution presented in Table~\ref{tab:0537_subset_solutions} using \textsc{tempo2}.  This way, every arrival time corresponded to a particular rotation of the NS, allowing for phase coherence throughout the full data set. This technique allows the characterization of shorter-term timing variations which may be missed when only fitting long-term solutions.

Within each inter-glitch data segment, a series of short-term timing solutions was fit using  rotation frequency $\nu$ and frequency derivative $\dot{\nu}$. Each solution comprised 10 TOAs, successively shifted forward in time by one TOA. The result of this partially coherent timing analysis is shown in Figure~\ref{fig:partial_coherent}, which presents the secular evolution of $\nu$ (plotted as the residual $\Delta\nu$, the difference from the best-fit rotation frequency published by M06, at reference epoch MJD 52061.3) and $\dot{\nu}$ over the entire \rxte{} data set.

We performed a linear fit to the overall trend in the short-term frequency derivative data, and from the fit slope calculated a second frequency derivative $\ddot{\nu} = (-8.2 \pm 0.3) \times 10^{-22}\,$s$^{-3}$.  From this, we determined a nominal value for the braking index of \psr{0537} to be $-1.28 \pm 0.04$, with uncertainties in both $\ddot{\nu}$ and $n$ derived using a bootstrap resampling method, and representing the $95\%$ error interval.  This best fit is shown in Figure~\ref{fig:partial_coherent} as a red dashed line, with the 95\% uncertainty interval represented by the red region. 
This is similar to previously reported values of braking index; however, we believe its utility is limited for physical interpretation, due to the high glitch activity in this pulsar and the difficulty it places on determining its intrinsic spin-down rate.    In \S\ref{sec:recovery_model}, we address this by determining a second frequency derivative that may be free from the effects of the glitch recovery, and use it to determine the braking index.  
We further argue in \S\ref{sec:braking_index} that the negative values of the braking index previously derived for this pulsar are an artifact of the glitch behavior.

\input{0537_glitch_table}

\section{Glitches in PSR~J0537$-$6910}
\label{sec:glitches}

\subsection{Glitch amplitude and activity rate}
\label{sec:glitch_size}

We extrapolated the ephemerides found in \S\ref{sec:timing} for each data segment in order to calculate rotation frequency $\nu$ and frequency derivative $\dot{\nu}$ at each preceding and subsequent glitch epoch. We then used these extrapolated values to determine the glitch magnitudes $\Delta\nu$ and $\Delta\dot{\nu}$ at each epoch.  These are summarized in Table~\ref{tab:0537_glitches}.

\begin{figure}
  \begin{center}
    \includegraphics[width=0.4\textwidth]{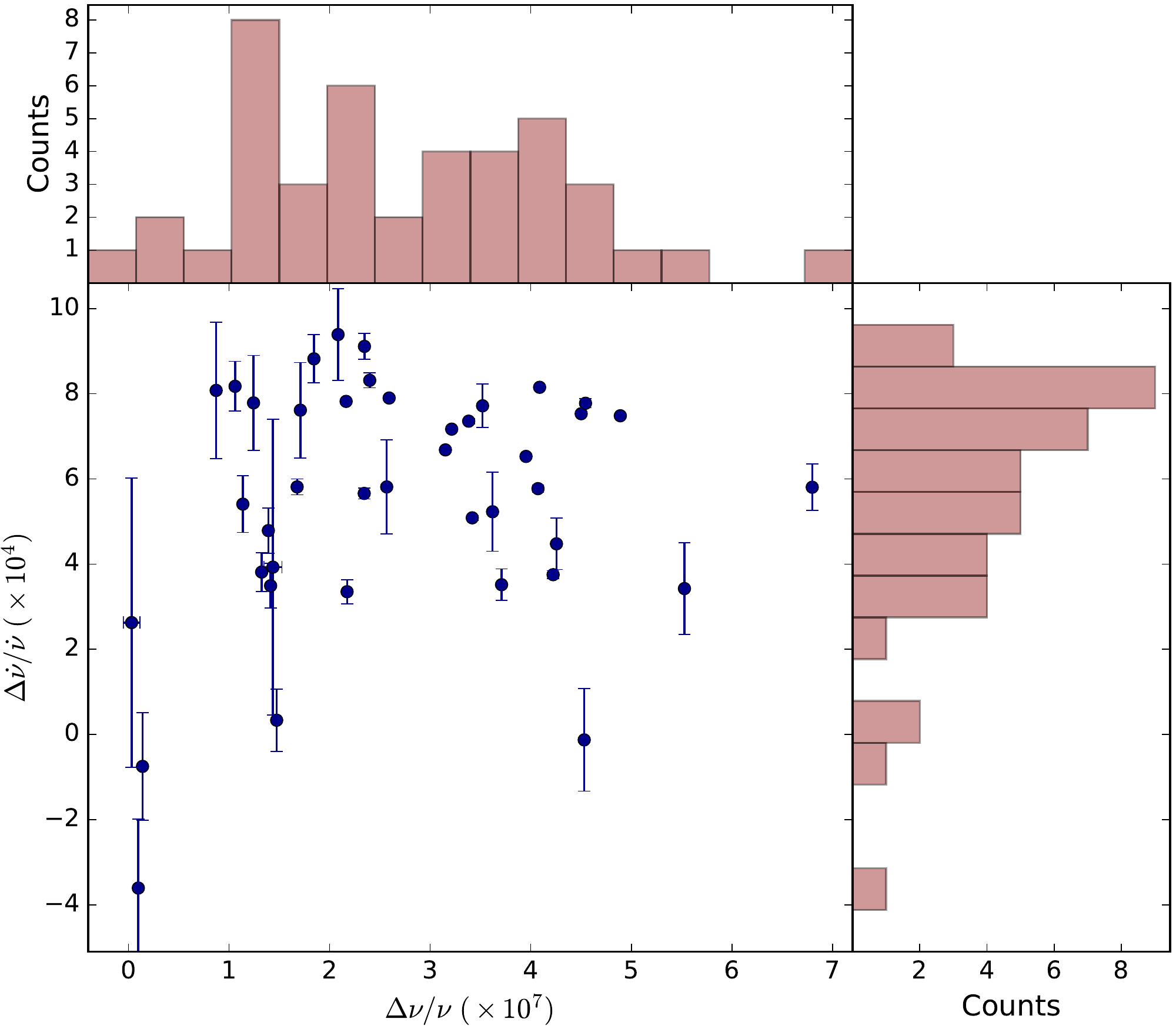}
    \caption{Glitch sizes in frequency ($\Delta\nu/\nu$) plotted against those in frequency derivative ($\Delta\dot{\nu}/\dot{\nu}$), along with corresponding 1-D histograms of each. 
\label{fig:glitch_size1}}
  \end{center}
\end{figure}

\begin{figure*}
  \begin{center}
    \includegraphics[width=0.98\textwidth]{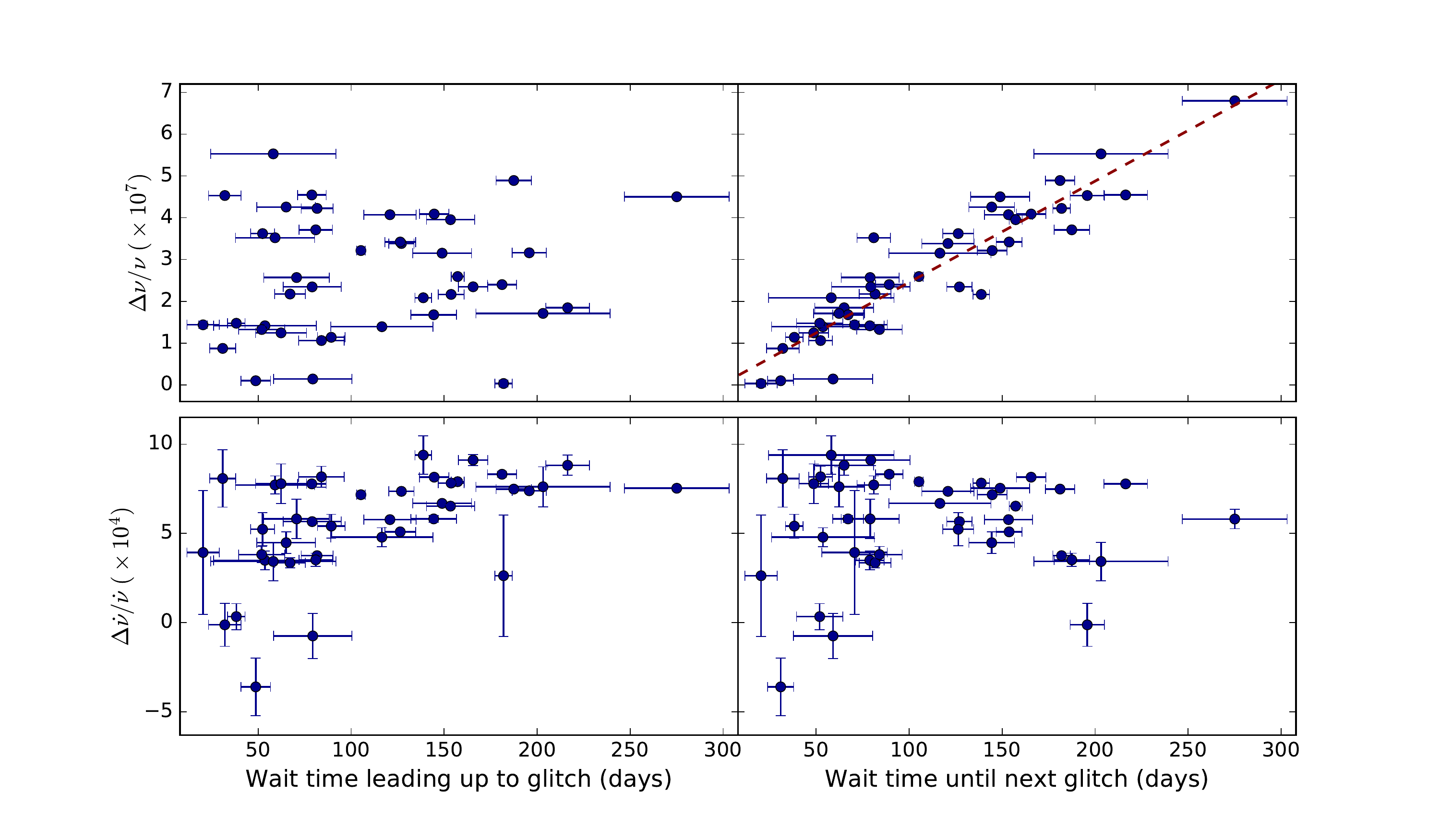}
    \caption{Fractional glitch magnitudes as a function of glitch wait times, with left panels in each row plotted against wait times prior to each glitch, and right panels showing wait times after each glitch. \textit{Top row}: Rotation frequency glitch sizes ($\Delta\nu/\nu$). There is a clear linear correlation with the amount of time elapsed before the next glitch (right), while there is no discernible correlation with the amount of time that has passed prior to the glitch (left). \textit{Bottom row}: Fractional frequency derivative glitch sizes ($\Delta\dot{\nu}/\dot{\nu}$).  In both cases, there is no apparent correlation, however, we do notice a preferred amplitude, supporting our claim in \S\ref{sec:recovery_model} of a common recovery behavior.  Uncertainties in $\Delta\nu/\nu$ (smaller than marker size) and $\Delta\dot{\nu}/\dot{\nu}$ are from those found in Table~\ref{tab:0537_glitches}; error bars for wait times correspond to the combined uncertainties in glitch epochs that precede and follow the corresponding inter-glitch data segment.\label{fig:glitch_size2}}
  \end{center}
\end{figure*}

We find no obvious relationship between the glitch size, $\Delta\nu$, and corresponding spin-down rate $\Delta\dot{\nu}$ (see Figure~\ref{fig:glitch_size1}).  
We also investigated the relationship between glitch sizes and corresponding pre- and post-glitch wait times; our results are shown in Figure \ref{fig:glitch_size2}.  
We find no correlation between the glitch magnitude in $\nu$ and corresponding wait time preceding the glitch.  However, we corroborate what was first reported by M06 to be a \emph{strong} correlation between glitch amplitude and the wait time until the following glitch, with slope $\sim0.2\,$d$^{-1}$.  As we discuss in more detail in \S\ref{sec:dicussion_glitch}, we believe this corresponds to a situation described by \citet{hm15}, where the initially random glitch occurrence and magnitude gives rise to periods of stability and build-up time proportional to the glitch size.
The glitches in $\dot{\nu}$ do not show a definitive correlation with wait time, however there does appear to be a preference for large changes in frequency derivative, particularly corresponding to longer wait times leading up the glitch.

It is clear that \psr{0537} is a particularly active young glitching pulsar. We can heuristically describe this activity with the so-called \emph{activity parameter}, which measures the accumulated frequency change in the pulsar over time.  This is given by $A_g = \sum{(\Delta\nu/\nu})/\Delta t$, where $\Delta t = t_f - t_i$ is the observing time span, beginning at time $t_i$ and finishing at time $t_f$.  This sum is taken over all detected glitches between $t_i$ and $t_f$.  For \psr{0537}, we find $A_g = 8.8 \times 10^{-7}\,$yr$^{-1}$ for the full data set.

\begin{figure}
  \begin{center}
    \includegraphics[width=0.5\textwidth]{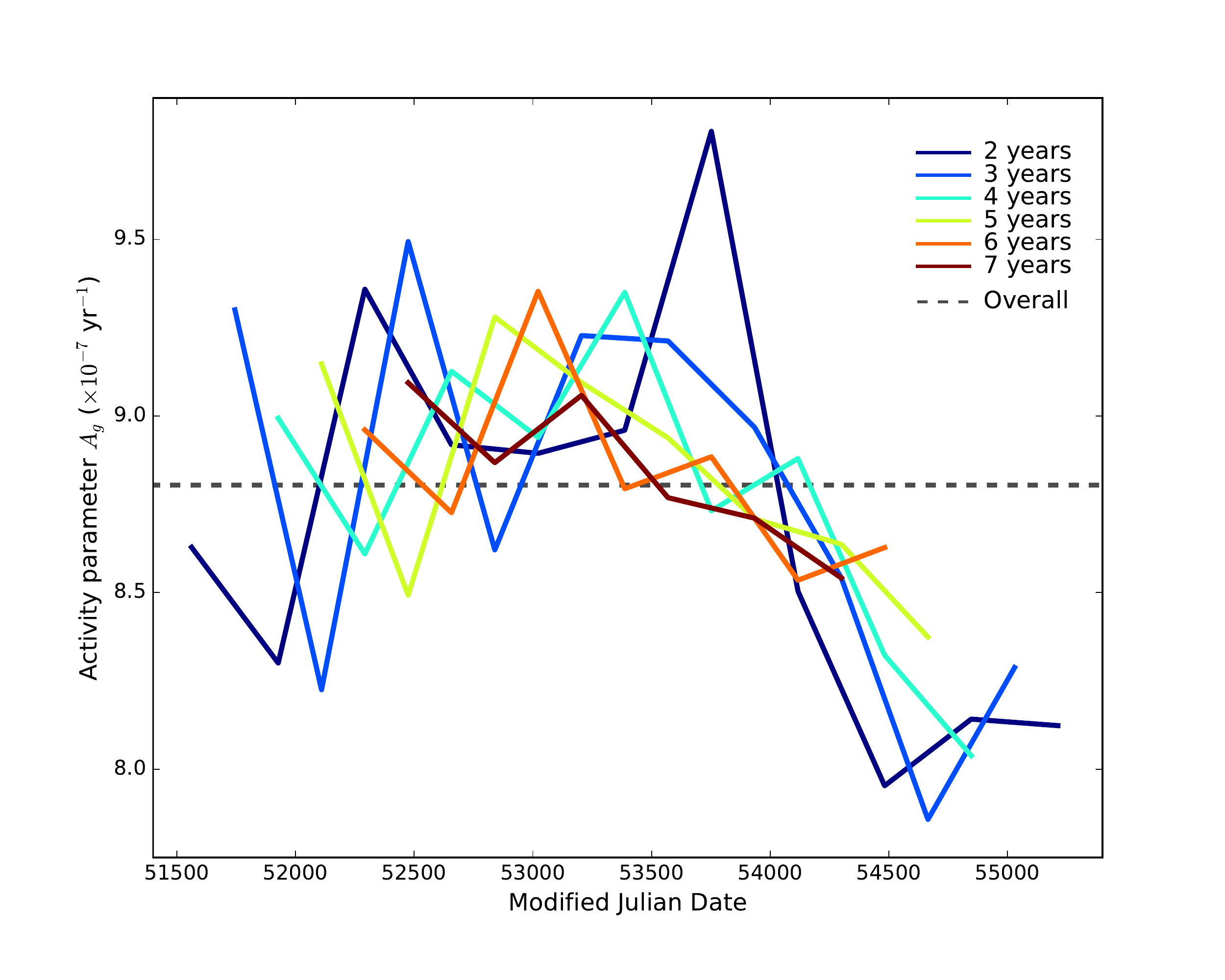}
    \caption{Activity parameter $A_g$ for PSR~J0537$-$6910 plotted over time, for several values of time span $\Delta t$ plotted in different colors. Lines connecting points for a given time span value were added as a guide to the eye. The overall value of $A_g=8.8 \times 10^{-7}\,$yr$^{-1}$, determined from the entire data set, is plotted as a horizontal dashed line.\label{fig:activity}}
  \end{center}
\end{figure}

The high glitch rate in \psr{0537} allows us for the first time to investigate the \emph{evolution} of a pulsar's activity. We determined $A_g$ using several values for $\Delta t$, ranging from $2-7\,$yr; for each, we incrementing the start time $t_i$ by one year, with the restriction that $t_f$ must fall within the data span. 
Figure~\ref{fig:activity} plots $A_g$ at the midpoint of each evaluated data span, for each span duration $\Delta t$, and suggests a possible systematic drop in the glitch activity near the end of our data set. However, this is not conclusive and requires confirmation with additional data.

\subsection{Recovery modeling}
\label{sec:recovery_model}

In an attempt to find an intrinsic braking index, we aimed to gain a better understanding of the $\dot{\nu}$ evolution seen in Figure~\ref{fig:partial_coherent}.  To do so, we performed a Markov-Chain Monte Carlo fit to those data in order to describe the $\dot{\nu}$ glitch recoveries. We included data segments in our fit having more than five measurements of the frequency derivative.  

We additionally found that several of the data segments showed a brief ``uptick'' and in some cases, a ``downtick'' in $\dot{\nu}$, of varying amplitudes and durations, typically occurring tens of days before a glitch. 
We principally observe this effect in the best sampled data segments. It is unclear from the current data set how frequently this occurs, and to what extent their rate of occurrence and apparent durations are biased by data sampling. Interestingly, this also means that we cannot rule out the existence of these upticks or downticks before every glitch, nor can we say very much about their general characteristics, or whether they represent typical pre-glitch behavior.
To avoid influence on our fit due to these currently unpredictable variations, we excluded data within 60 days prior to each glitch.  This ensures that the recovery model fits are based on the glitch depth measurements in Table~\ref{tab:0537_glitches} found from extrapolations of the coherent inter-glitch timing solutions.  This pre-glitch behavior remains unexplained, and is beyond the scope of this work.  However, it warrants further detailed investigation, as it may provide unique insight into the glitch mechanism itself.

We found that a single dual-component model comprising a linear function and a decaying exponential can reproduce the post-glitch behavior for all fit $\dot{\nu}$ recoveries.  Specifically, we maximized the likelihood of the following empirical model compared to the observed $\dot{\nu}$ as a function of time $t$:
\begin{equation}
\label{eqn:recovery}
\dot{\nu}(t-t_{0,i}) - \dot{\nu}_{0,i} = m(t-t_{0,i}) + A\left(1 - e^{\frac{-(t-t_{0,i})}{\tau}}\right)\,,
\end{equation}
where $t_{0,i}$ represents the $i$th observed glitch epoch.  In performing this MCMC fit we let the following vary: $A$ and $\tau$, the amplitude and decay timescale of the exponential component, respectively; the slope $m$ of the linear component; and $\dot{\nu}_{0,i}$, the $i$th post-glitch value of each frequency derivative, at time $t_{0,i}$.  We chose conservative uniform prior distributions for these parameters, based on the extreme values measured for $\dot{\nu}$ and glitch amplitudes, and we used preliminary least-square fit values to these parameters as starting estimates.  

\begin{figure}
  \begin{center}
    \includegraphics[width=0.5\textwidth]{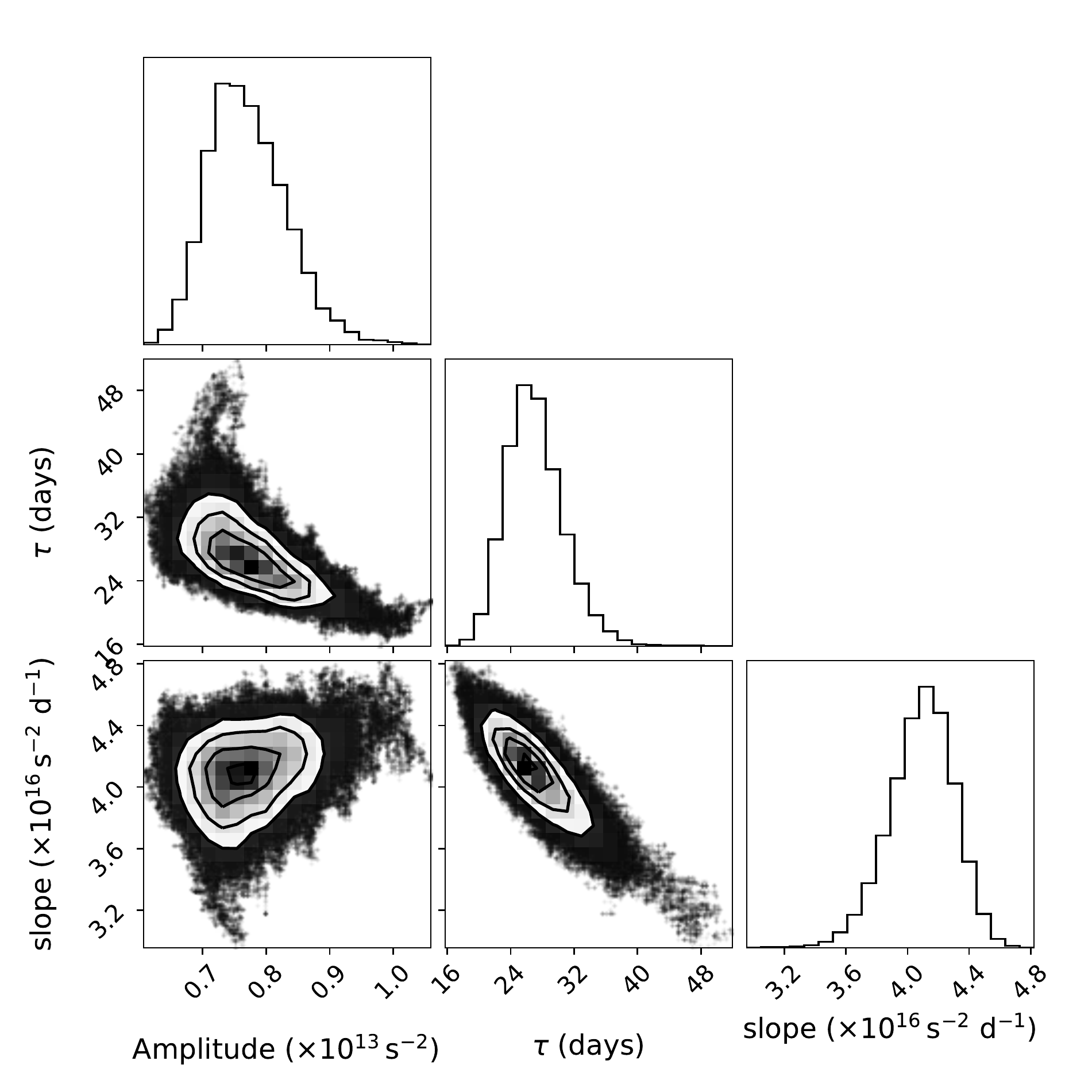}
    \caption{Summary of MCMC results for simultaneous fitting of $\dot{\nu}$ data. Shown are 2-D PDFs for exponential component amplitude $A$ and decay timescale $\tau$, as well as linear component slope $m$, with marginalized 1-D PDFs on the top row of each column \citep[made with the \texttt{corner.py} plotting routine;][]{corner}.\label{fig:mcmc_fit}}
  \end{center}
\end{figure}

\begin{figure}
  \begin{center}
    \includegraphics[width=0.5\textwidth]{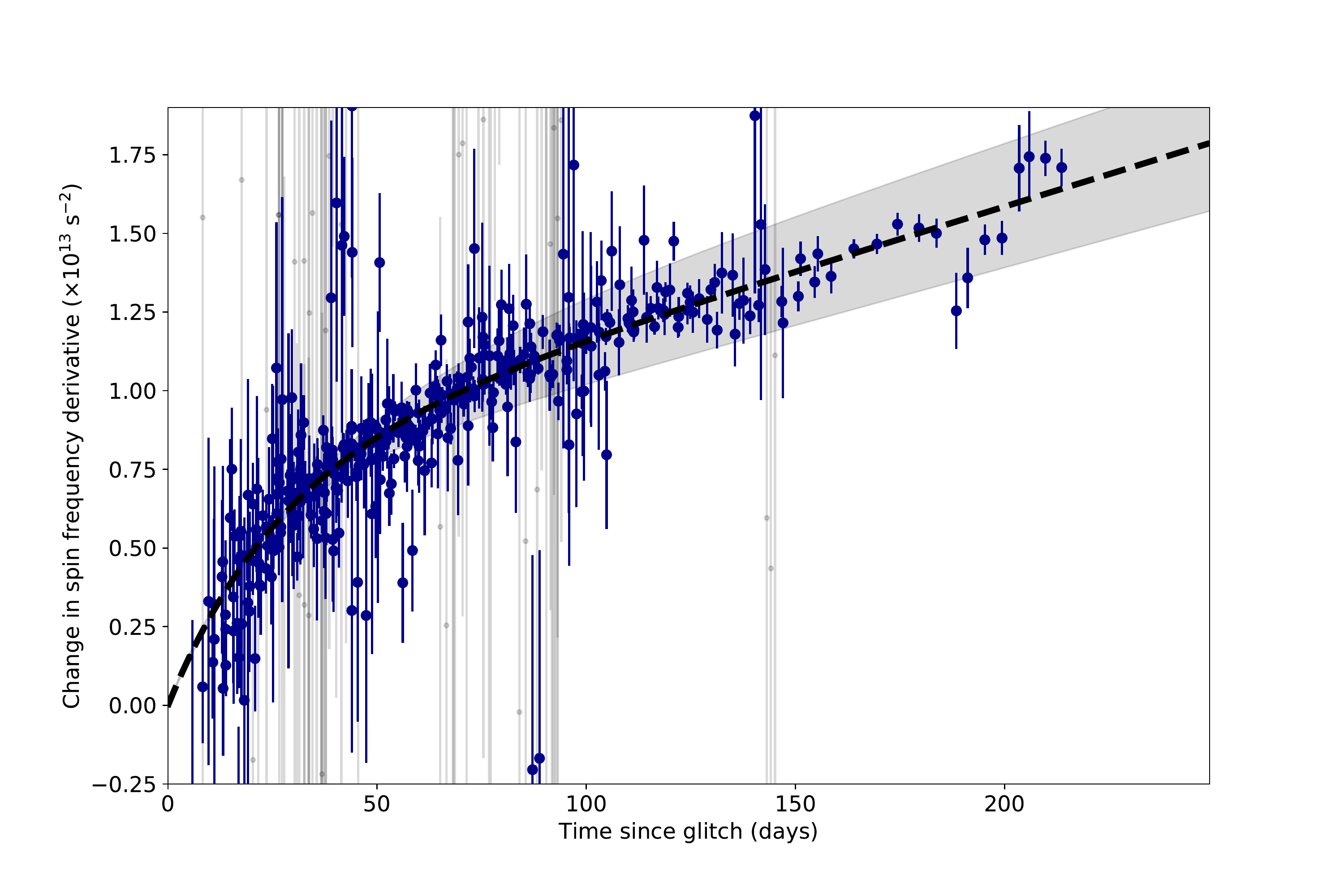}
    \caption{Rotation frequency derivative $\dot{\nu}$ plotted against time since glitch.  Values from all inter-glitch intervals included in the MCMC fit are plotted in blue, with the immediate post-glitch spin-down rate $\dot{\nu_0}$ subtracted from each. Faded grey points are those with errors larger than $7.5\times 10^{-14}\,$s$^{-2}$ (and therefore contributed little weight in the fit). Overplotted as a black dashed curve is the best-fit exponential-plus-linear model curve determined from median values of fit parameters from the MCMC fit, as well as the corresponding 95\% highest posterior density region.\label{fig:f1_stack_fit}}
  \end{center}
\end{figure}

Posterior probabilities for these parameters and marginalized PDFs are shown in Figure~\ref{fig:mcmc_fit}. We find all glitch recoveries are well fit by a linear slope $m=(4.1 \pm 0.4)\times 10^{-16}\,$s$^{-2}\,$d$^{-1}$, exponential amplitude $A=(7.6 \pm 1.0) \times 10^{-14}\,$s$^{-2}$, and decay timescale $\tau=27^{+7}_{-6}\,$d.  Stacked relative inter-glitch frequency derivative values $\dot{\nu} - \dot{\nu}_{0,i}$, plotted against time since the glitch $t-t_{0,i}$, are shown in Figure~\ref{fig:f1_stack_fit}, along with the above best fit for the $\dot{\nu}$ recovery and 95\% highest posterior density region. 

The slope of the linear component corresponds to a second frequency derivative $\ddot{\nu} = (4.7\pm 0.5) \times 10^{-22}\,$s$^{-3}$ that remains approximately constant throughout the data set. One interpretation of this may be an intrinsic $\ddot{\nu}$ for \psr{0537}, without contamination from the effects of glitch recovery (see \S3.1).  In this case, we find an overall braking index of $n=7.4 \pm 0.8$ based on this value.
While acknowledging that we may not have been able to fully account for all sources of systematic error in the quoted uncertainties, our result is inconsistent with a negative braking index for \psr{0537}.
We have additionally performed a consistency check by omitting data from the second---and longest---inter-glitch interval, which clearly dominates the linear term in the fit, and repeating our analysis. We find $n=10\pm1.1$, agrees within $2\sigma$ with our value for the global braking index.
In this interpretation, \psr{0537} has the largest long-term braking index known; the physics behind this large value has no current explanation, and is possibly due to a different cause than in other pulsars.
Discussion of the braking index and the repeated recovery behavior of \psr{0537} are found in \S\ref{sec:braking_index}.

\section{Radiative changes} 
\label{sec:rad_change}

\subsection{Search for flux variations}

We conducted a flux analysis of our data set to search for flux variations associated with the observed glitches.   
To do so, we first filtered our event data to exclude known high-background epochs that may have biased our flux measurements.  This included the rejection of data during South Atlantic Anomaly passage, Earth occultation and bright Earth effects, and/or electron contamination.  We also rejected epochs with pointing offsets greater than $0.02\degrees$. 
As well, when checking for long-term flux changes we only consider data taken with PCU2, as PCUs 0 and 1 lost their propane layers during this monitoring campaign \citep[see, e.g.,][as well as the {\it RXTE} guest observer online facility\footnote{\url{http://heasarc.gsfc.nasa.gov/docs/xte/}}]{jmr+06, chp+10}. 

We determined the pulsed count rate with a method that uses the RMS variations in pulsed flux \citep[see][for more detailed descriptions]{ahk+13,abp+15}.
This method is more robust than the traditional flux estimation via integrating the pulse profile
above the background level because the latter has a large bias due to uncertainty in the profile minimum.

We searched for pulsed flux variation in two energy bands: 2--5\,keV, to be sensitive to changes with thermal signatures as predicted in some starquake models (see \S\ref{sec:discussion}); and the 2--30\,keV band, in order to maximize our sensitivity. 
We find that there are no significant pulsed flux enhancements in any individual observation with a constant-flux fit giving a reduced $\chi^2_\nu$ of 0.93 and 1.12 in the 2--5\,keV and 2--30\,keV bands, respectively.
We place a 99\% confidence limit on any pulsed flux change of 120\% in the 2--5\,keV band, and 105\% in the 2--30\,keV band.

\begin{figure}
  \begin{center}
    \includegraphics[width=0.5\textwidth]{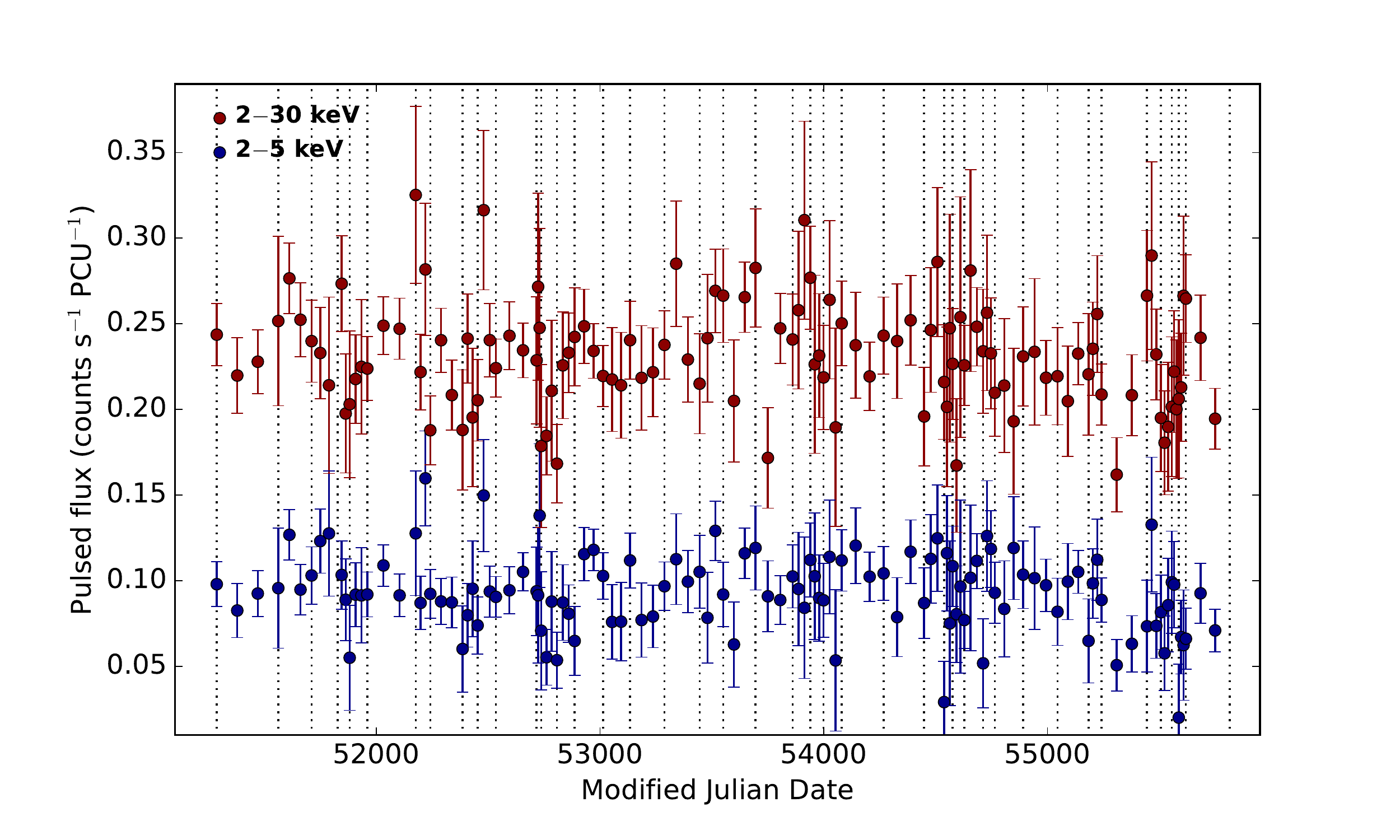}
    \caption{RMS pulsed flux of PSR\,J0537$-$6910 over the {\it RXTE} campaign. The blue points show the pulsed flux in the 2--5\,keV band, and the red points in the 2--30\,keV band. Glitch epochs are indicated by dotted vertical lines.\label{fig:flux}}
  \end{center}
\end{figure}

We then combined observations to look for lower-level pulsed flux variations.
To do so, we divided each inter-glitch period into three equal sections.
In each of these sections, we folded all observations with the local timing solution, and combined aligned profiles using the ML method described in \S\ref{sec:timing}. 
Again, as shown in Figure~\ref{fig:flux}, no pulsed flux enhancements were detected, with a constant-flux fit giving a reduced $\chi^2_\nu$ of 1.03 and 0.93 in the 2--5\,keV and 2--30\,keV bands, respectively.
We place a 99\% confidence limit of  49\% for any change in the 2--5\,keV band and 26\% in the 2--30\,keV band.

\subsection{Search for profile shape changes}
In order to test for changes in profile morphology, we used the template profile we created for our timing analysis described in \S\ref{sec:timing} as a model for the profile shape of \psr{0537}. 
For each observation we folded the both the 2--5\,keV and 3--20\,keV photons using the local timing solution into respective 32-bin profiles. 
We then scaled the template to match the profile after removing the mean values, in order to account for any relative background and/or exposure length differences.
We then aligned each profile in pulse phase with the template, using the ML technique.

We determined the goodness-of-fit of our template profile to each aligned and scaled profile by subtracting the two and calculating the $\chi^2$ statistic for the residuals.  We find consistency with the expected $\chi^2$ distribution in both energy bands,
including for profiles at epochs surrounding glitches.
 
We therefore detect no significant change in profile shape over the \rxte{} observing campaign, including at or near glitch epochs.

\subsection{Search for burst activity}

We conducted a search for short bursts from \psr{0537} over the nearly 13-yr \rxte{} data set, using the method presented in \cite{gkw04} and
\cite{sk11}.
We created the time series from the event data using $0.01$-s and $0.1$-s time bins, in the $3-20\,$keV energy range. These timescales were chosen since this represents the typical duration of magnetar-like bursts observed in multiple sources \citep[see e.g.][]{ckh+15}.
We did this over overlapping 50-s time intervals, to mitigate effects from the variable background rate from \rxte{}.
We searched for time bins with more counts than would be expected from a Poisson distribution centered at the mean of that observation.
In order for a candidate flagged in this manner to be considered real, we required that it be seen in all active PCUs.
We found no significant bursts at the $0.01$--$0.1$-s timescales in this data set.

\section{Discussion}
\label{sec:discussion}

Although glitches have now been observed in dozens of pulsars, the timing behavior of \psr{0537} is exceptional. Not only does it have the highest known activity parameter, but the time evolution of this parameter can be extracted. The waiting time for a glitch to occur has a remarkably strong observed correlation with the size of the glitch that precedes it. Finally, the recovery of its glitches can be described using a single model consisting of an exponentially decaying term and a linear term.  The latter can be interpreted as a long-term intrinsic second rotation frequency derivative, from which we can derive a braking index. Unlike in \psr{0537}, glitch sizes in the vast majority of known pulsars vary over several orders of magnitude \citep{elsk11} and show no periodicity.  The majority of pulsar glitch models have therefore focused on reproducing this behavior \citep{hm15}. In what follows, we discuss the implications of these observations for existing pulsar glitch theory. 

\subsection{Glitch sizes, recovery and waiting times}
\label{sec:dicussion_glitch}

Glitches are believed to be the result of the impulsive exchange of angular momentum between the superfluid content of the pulsar and its normal matter \citep{bppr69, ai75}. In the standard picture, non-superfluid material spins down steadily due to its strong coupling to the torque exerted by its magnetic field, whereas the superfluid content lags behind. Once this increasing lag exceeds a critical value, angular momentum is transferred to the normal material, leading to a sudden spin-up glitch. In this scenario, the critical lag is much larger than the lag developing between glitches \citep{alp77}, therefore no correlation between glitch size and waiting time is expected if the glitch sizes span over a few orders of magnitude, as is most often the case. However, as most glitches in \psr{0537} have similar amplitudes, we expect the critical lag to be close to the lag that develops during the inter-glitch interval. 

We can evaluate the critical lag using the activity parameter $A_g$ found in our analysis. To do so, we first estimate the moment of inertia $I_{sf}$ of the superfluid participating in the glitch as a fraction of the total moment of inertia $I$ \citep{lel99}, noting that this may be an underestimate by a factor of a few, if  the superfluid is entrained by the crustal lattice \citep{cha13, agh13}:
\begin{eqnarray}
\frac{I_{sf}}{I}= A_{g}\nu/\dot{\nu}=8.7\times 10^{-3}\,.
\label{MOM_INERTIA}
\end{eqnarray}
We then estimate the critical lag $\delta \nu$ that accumulates between the crustal matter and the superfluid, and triggers the largest observed glitch, $\Delta \nu= 4.2\times 10^{-5}$s$^{-1}$. Assuming that the normal and superfluid materials synchronize completely after the glitch, we can then write an expression for angular momentum conservation during a glitch:
\begin{eqnarray}
(I-I_{sf})\nu + I_{sf}(\nu+\delta \nu) = I (\nu + \Delta \nu)\,.
\label{ANG_MOM}
\end{eqnarray}
Solving for the critical lag, we find $\delta \nu \approx 4 \times 10^{-3}$~s$^{-1}$. 

There are two scenarios among glitch models that predict a correlation between glitch magnitudes and waiting times: the crust-quake model, with and without the inclusion of superfluidity \citep{rud76, accp94, bp71}; and the snowplough model \citep{piz11}. 
In the crust-quake scenario, the pulsar is approximated by an oblate spheroid because of rapid rotation. As it spins down, there is a decrease in the centrifugal forces that sustain the equilibrium of the crust.  Consequently, increased stress is exerted onto the lattice. Eventually, the crust can no longer support this extra stress and yields, relaxing to a state with lower eccentricity and thus moment of inertia. Due to angular momentum conservation, this change in the moment of inertia is accompanied by a sudden increase in the angular velocity---a glitch. One of the principal predictions of this model is a correlation between the size of a glitch and the waiting time until the next glitch. This is because bigger glitches will relieve more stress, and therefore more time will be required for the pulsar to once again reach a critical point, producing the next glitch. The waiting time is given by:
\begin{eqnarray}
t_{w}=\frac{4 \pi A_{0}^2\tau_{c}}{A_{1} I \nu{^2}}\frac{ \Delta{\nu}}{ \nu}\,,
\label{T_wait}
\end{eqnarray}
where $A_{0}=\frac{3}{25}\frac{GM^{2}}{R}$ and $A_{1}=\frac{57 \mu}{50}\frac{R^{3}}{3}$, where $G$ is Newton's constant, and we assume that the moment of inertia of the pulsar $I=1.5\times 10^{45}$g~cm$^{2}$. $\tau_{c}$ is the characteristic age of the pulsar, $\mu\approx 10^{30}$g cm$^{-1}$s$^{-2}$ is the shear modulus \citep{svo+91}, and we assume that $M=1.4M_{\odot}$ and $R=1.4\times 10^{6}$cm is its mass and radius, respectively. Substituting the above values in Equation~\ref{T_wait} we find that for a $\Delta \nu/ \nu = 10^{-7}$ glitch (as is typical for \psr{0537}), $t_{w}\approx 10^{3}$~yr, three orders of magnitude higher than is observed. If the superfluid component is included in the crust-quake model \citep{rud76, accp94}, the waiting time can decrease substantially. In this case, the interaction between the crust and the superfluid content triggers a crust quake and an accompanying glitch. The critical lag to trigger a glitch is $\delta \nu =  \sigma_{crit} \frac{\mu}{p}\frac{GM}{2 \pi R^{3} \Omega}$ rad~s$^{-1}$, where the critical strain $\sigma_{crit}\approx 0.1$ \citep{hor10} and $\mu/p\approx 10^{-2}$ at the base of the crust \citep{svo+91}. Using these quantities we find that the critical lag between the crust and the superfluid to trigger a glitch through a crust quake is $\delta \nu \approx 0.3$~s$^{-1}$, which is 100 times higher than the lag that can develop between two glitches, found from conservation of angular momentum (Eq.~6). Unless the critical strain were at least two orders of magnitude lower, these events seem unlikely to be due to crust quakes. Prior to the realization that $\sigma_{crit}=0.1$, this scenario had been explored in M06 using a lower value for this parameter, based on analogies to terrestrial materials. 

A further prediction of the crust-quake models is a possible enhancement of the X-ray flux, caused by the conversion of kinetic energy into heat \citep{ts01, fle00}. Taking into account angular momentum conservation, the kinetic energy decrease during a glitch is: $|\Delta E_{K}|=E_{k,b}-E_{k,a}$, where $E_{k,b}=2\pi^{2}(I_{c}\nu^{2}+I_{sf}(\nu +\delta \nu)^{2})$, $E_{k,a}=2\pi^{2}(I_{c}+I_{sf})(\nu+\Delta \nu)^{2}$, and $I_{c}=I-I_{sf}$. Substituting in the values derived from Equations~\ref{MOM_INERTIA} and \ref{ANG_MOM} we find that $|\Delta E_{k}|=1.5\times 10^{42}\,$erg. The relative enhancement of the X-ray flux depends on the core temperature.  For example, for a core temperature of $10^{8}\,$K an increase by at least $2\%$ is expected to occur a few days after the glitch, whereas a core temperature of $10^{7}\,$K can result in an enhancement of up to an order of magnitude \citep{ts01}. The latter can be ruled out by the X-ray observations presented in this work (see \S5.1).

Persistent changes in spin frequency derivative originating from glitches may reflect a reorientation of the magnetic axis with respect to the rotation axis. If this is the case, we may expect a change in the pulsed profile of $\sim 1\%$ for a  $\Delta\dot{\nu}/\dot{\nu}\sim 10^{-3}$ \citep{le97}. Neither the \rxte{} data set used for this work, nor other current X-ray telescopes have the sensitivity capable to detect this level of change, but this may become feasible with future proposed missions such as ESA's proposed large area X-ray mission {\it ATHENA}\footnote{http://www.the-athena-x-ray-observatory.eu/}. 

In the ``snowplough'' model \citep{piz11}, the forces exerted by the superfluid vortices onto the crust are considerably weaker than those of the canonical superfluid glitch model \citep{alp77}, as no normal matter layer is present between the inner crust and the core. Therefore, once a glitch occurs, a much larger fraction of the angular momentum stored in the superfluid is released to the crust. This leads to the depletion of the angular momentum reservoir. Following a glitch, some time is then required to accumulate lag that will result in the release of stress during the next glitch. According to this model, the fractional jump in frequency derivative is given by:
\begin{eqnarray}
\frac{\Delta \dot{\nu}_{\rm a}}{\dot{\nu}}= \frac{Q(1-Y_{gl})}{1-Q(1-Y_{gl})}\,,
\end{eqnarray}
where $Q=0.95$ is the neutron fraction of the star and $Y_{gl}$ is the fraction of vorticity coupled to the crust, a parameter that is assumed in the model to be approximately $10\%$ and $\Delta \dot{\nu}_{\rm a}$ is the change in frequency derivative immediately after the glitch. This would imply a $\Delta \dot{\nu}_a/\dot{\nu} \approx  6$, which should decay within a few minutes after the glitch \citep{Haskell:2012}. However, our data have insufficiently high time resolution to confirm the applicability of this scenario.

The correlation between glitch sizes and waiting time to the next glitch strongly implies that a considerable fraction of the lag developed between normal and superfluid matter during an inter-glitch interval is depleted during a glitch. 
Moreover, the fact that all glitch recoveries in this pulsar can be described by approximately the same model implies that the same process is involved in the recoupling between normal and superfluid material following each glitch. This has not been observed in other young pulsars.  Even the Vela pulsar, which has qualitative similarities to \psr{0537}, does not demonstrate the same strong correlation between glitch size and inter-glitch time, suggesting that the lag reservoir is not fully depleted during a glitch.

\subsection{Braking index}
\label{sec:braking_index}
In contrast to previous work, we argue against a negative intrinsic braking index for \psr{0537} \citep[see M06;][]{els16}, measured by fitting a second frequency derivative to the long-term trend in the $\dot{\nu}$ evolution.
In this way, we find what we propose to be a true braking index closer to $n \approx 7$, measured by effectively subtracting the (common) exponential decay component of the glitch recoveries from the inter-glitch $\dot{\nu}$ evolution, then taking the slope of the residual linear component to be the uncontaminated second spin derivative.
The former negative values posed a challenge to theoretical models of pulsar spin-down which thus far can reproduce values only in the range $1\leq n \leq 5$.
Although our measured value of $\approx 7$ remains out of this range, it can be accommodated in the dipole spin-down model by assuming evolution in either the magnetic field or the moment of inertia over time.

The electromagnetic torque acting on the pulsar is 
\begin{eqnarray}
\frac{d \left(I \nu\right)}{dt}= -k B_{d}^2 \nu^{3}\,,
\label{EDOT}
\end{eqnarray}
where $k$ is a numerical parameter that depends on the radius of the neutron star and the angle between the rotation and magnetic axes, and $B_{d}$ is the intensity of the magnetic dipole \citep{spi06}. If the magnetic field evolves with time while the moment of inertia remains constant, then from equation (\ref{EDOT}) we can express the braking index as $n=3-4\tau_{c}\dot{B}_{d}/B_{d}$; for $n=7.4$ we obtain a decay timescale for the magnetic field $|B/\dot{B}|=4.5\,$kyr, which is  faster than the expected decay rate due to Ohmic dissipation of the magnetic field  but may be achieved if the decay is accelerated by the Hall effect with an appropriate toroidal field \citep{gc15, gwh16}.

A time-varying moment of inertia can affect the braking index. Differentiating Equation~\ref{EDOT}, while keeping the magnetic field constant and allowing for the moment of inertia to vary, we find $n=3-2\tau_{c} \dot{I}/I - 4\tau_{c}^{2}\ddot{I}/I$. This scenario has been explored in models where normal matter turns into superfluid as pulsar ages \citep{ha12}. However, it is not compatible with the current measurement of the braking index, as the values for $\dot{I}/I$ and $\ddot{I}/I$ obtained by their model lead to $n<3$. Alternatively, changes in the obliquity angle can lead to a braking index of $\sim7$ if the rotational and magnetic axes become more aligned, at a rate of approximately $1^{\degrees}$ per century for an obliquity of $30^{\degrees}$.  
This could in principle be observed as a secular change in pulse profile shape. However, we do not observe this in the current data set, which is not adequately sensitive to perceive such changes; at the above rate of $1^{\degrees}$ per century, this is estimated to be at the $\sim1\%$ level over the entire data span.
We remark that a braking index $n=5$ is expected if the pulsar spins-down exclusively due to gravitational wave radiation. If we assume that the spin-down is caused by the combined effect of magnetic braking and the emission of gravitational waves \citep{Palomba:2000}, the decay rate of the magnetic field required to explain this behavior will have to be even faster than the $4.5\,$kyr mentioned above, making such an explanation even more demanding.

We can additionally quantify the effect of glitches on the evolution $\dot{\nu}$. Their impact comes through two basic routes: (a) the change of $\Delta\dot{\nu}$ that occurs during the glitch itself, and (b) the glitch recovery process. In a similar manner to the activity parameter, we can define a second activity parameter 
\begin{eqnarray}
A^{\prime}_{g}\equiv\frac{1}{\tau_s}\sum_i\frac{\Delta \dot{\nu}_{i}}{\dot \nu_{i}}\,,
\end{eqnarray}
which expresses the time-averaged effect of glitches on the rotation frequency derivative. 
We find that for \psr{0537} $A^{\prime}_{g}=1.9\times 10^{-3}$ yr$^{-1}$, which leads to a change in the period derivative by $\langle d\dot{\nu}/d t\rangle= -1.2\times 10^{-20}\,$~s$^{-2}$, and an average glitch size $\langle \Delta \dot{\nu}\rangle=1.1\times 10^{-13}\,$s$^{-2}$. The overall trend in $\dot{\nu}$ shown in the lower panel of Figure 2 illustrates the effect of this high glitch activity: while the spin-down rate has a piecewise increasing trend, it decrease during glitches by an amount that eventually dominates its long-term evolution. 

We have modeled the glitch recovery as an exponential in order to remove it from the intrinsic $\dot{\nu}$ evolution. Using our best-fit parameters, we find that the spin frequency derivative will recover by $A=7.6\times 10^{-14}\,$s$^{-2}$, which is smaller than the average glitch size. Therefore, even if sufficient time is given for the pulsar to recover---generally found to be the case for \psr{0537}, as the average time between glitches is $\sim 100\,$d, about four times the recovery timescale of $\tau=27\,$d---the net effect will be a decreasing $\dot{\nu}$. 
A similar result has been recently reported for Vela, for which it has been proposed that the intrinsic braking index is $n=2.81$ \citep{aabp17}.  This was done by using the timing solution just prior to each glitch, so that the effect of recovery on $\Delta \dot{\nu}$ is minimized. 

We therefore argue that the negative braking index quoted for \psr{0537} is not indicative of any extraordinary magnetospheric evolution, but is rather an artifact of a high rate of large glitches and their recoveries.  However, this leaves open the question of why and how \psr{0537} glitches so frequently.

\section{Conclusions}
\label{sec:conclusions}
We have analyzed 13 years of \psr{0537} observations taken by \rxte{}.  We have used a novel maximum-likelihood technique to search for pulsar rotation periods in both the reprocessed M06 data set and subsequent observations, until the decommissioning of \rxte{}. Our results broadly confirm the findings of M06.\footnote{After the submission of this article, we became aware of a parallel analysis of the same dataset by \cite{aeka18}. Our measurements of glitches and other timing parameters are broadly consistent.} Specifically, we have found a strong correlation between glitch magnitude and wait time until the following glitch, which we attribute to the almost complete depletion of the lag between the normal and the superfluid content of the star.  
We also found that a single model with one linear and one decaying exponential component does well to fit the recovery of \emph{all} glitches for which there was adequate data sampling.  The linear slope and decay timescale found are not easily reproducible with crust-quake and snowplough models.  Further, our analysis does not support the reality of a negative braking index; rather, we argue
it is a consequence of the pulsar's high glitch rate inhibiting standard spin-down.
In contrast, our recovery model implies a consistent value for the second rotation frequency derivative across the data set, corresponding to a positive braking index $n = 7.4 \pm 0.8$. 
This analysis presents an interesting problem for glitch physics and magnetic field evolution in young pulsars. Further observation of \psr{0537} with high time resolution X-ray telescopes such as the recently launched \textit{Neutron Star Interior Composition Explorer} \citep[\textit{NICER};][]{agb+14} will be valuable for investigating the extreme physics that governs the evolution of this and other young pulsars.

\acknowledgements
We thank A.~Cumming, D.~Antonopoulou, C.~M.~Espinoza, and N.~Andersson for useful discussions. R.F.A. acknowledges support from an NSERC Alexander Graham Bell Canada Graduate Scholarship.  K.N.G. acknowledges STFC grant ST/N000676/1 for support. V.M.K. receives support from an NSERC Discovery Grant and Accelerator Supplement, Centre de Recherche en Astrophysique du Qu\'{e}bec, an R.~Howard Webster Foundation Fellowship from the Canadian Institute for Advanced Study, the Canada Research Chairs Program and the Lorne Trottier Chair in Astrophysics and Cosmology. 
This research has made use of data and software provided by the High Energy Astrophysics Science Archive Research Center (HEASARC), which is a service of the Astrophysics Science Division at NASA/GSFC and the High Energy Astrophysics Division of the Smithsonian Astrophysical Observatory.



\end{document}

%% file: 0537_par_table.tex
\startlongtable
\begin{deluxetable*}{lccclll}
\tablecolumns{7}
\tablecaption{Individual coherent solutions for each data subset.\label{tab:0537_subset_solutions}}
\tablewidth{0pc}
\tablehead{
    \colhead{Subset}          &  
    \colhead{Reference epoch}  &  \colhead{Start epoch}  &  \colhead{Finish epoch}  &  
    \colhead{$\nu$}  &   \colhead{$\dot{\nu}$} &   \colhead{$\ddot{\nu}$} \\
    \colhead{No.}          &  
    \colhead{(MJD)}  &  \colhead{(MJD)}  &  \colhead{(MJD)}  &  
    \colhead{(s$^{-1}$)}  &   \colhead{($\times 10^{-10}$\,s$^{-2}$)} &   \colhead{($10^{-20}$\,s$^{-3}$)}     
}
\startdata 
1      &  51236  &   51197.1  &  51262.7  &  62.040263270(3)    &  -1.99219(3)    &   1.2(4)    \\
2      &  51428  &   51310.5  &  51546.7  &  62.0370003428(6)   &  -1.9922319(8)  &   0.487(5)  \\
3      &  51640  &   51576.6  &  51705.2  &  62.0333789952(11)  &  -1.992671(3)   &   0.74(3)   \\
4      &  51758  &   51715.9  &  51800.2  &  62.0313668785(16)  &  -1.992940(6)   &   1.49(9)   \\
5\tablenotemark{a}  &  51864  &  51854.1  &  51874.7  &  62.0295505393(19)  &  -1.99300(10)   &   \ldots   \\ 
6      &  51920  &   51886.9  &  51955.0  &  62.028594935(3)    &  -1.992772(11)  &   2.7(2)    \\
7      &  52064  &   51964.3  &  52165.3  &  62.0261437020(11)  &  -1.9927296(16) &   0.724(12) \\
8      &  52208  &   52186.8  &  52229.6  &  62.023675996(5)    &  -1.99320(2)    &   2.1(1.0)  \\
9      &  52317  &   52252.7  &  52381.5  &  62.0218253224(12)  &  -1.993040(3)   &   0.67(3)   \\
10     &  52417  &   52389.5  &  52445.5  &  62.020113712(4)    &  -1.99334(2)    &   1.7(5)    \\
11     &  52494  &   52459.9  &  52529.8  &  62.018801126(3)    &  -1.993273(11)  &   0.6(2)    \\
12     &  52636  &   52539.0  &  52715.4  &  62.0163816726(7)   &  -1.9931349(15) &   0.76(11)  \\ 
13\tablenotemark{a}  &  52636\tablenotemark{b}  &   52717.3  &  52728.1  &  62.0163820(5)     &  -1.9931(7)   &   \ldots   \\ 
14     &  52768  &   52745.4  &  52792.1  &  62.014117653(6)    &  -1.99366(5)    &   0.9(1.2)  \\
15     &  52853  &   52822.8  &  52883.8  &  62.012669324(4)    &  -1.993659(16)  &   2.1(4)    \\
16     &  52948  &   52889.2  &  53007.3  &  62.0110474794(11)  &  -1.993430(4)   &   1.39(4)   \\
17     &  53070  &   53019.7  &  53121.8  &  62.0089672274(19)  &  -1.993640(7)   &   0.96(8)   \\
18     &  53215  &   53146.9  &  53284.8  &  62.0064947565(16)  &  -1.993705(3)   &   0.79(3)   \\
19     &  53367  &   53290.9  &  53443.4  &  62.0039008356(19)  &  -1.993760(4)   &   1.10(3)   \\
20     &  53497  &   53446.8  &  53548.8  &  62.001677534(2)    &  -1.993959(5)   &   1.42(8)   \\
21     &  53619  &   53551.9  &  53687.2  &  61.999595598(2)    &  -1.993922(5)   &   1.37(5)   \\
22     &  53781  &   53702.8  &  53859.2  &  61.9968301021(19)  &  -1.993969(4)   &   0.91(4)   \\
23     &  53890  &   53862.0  &  53919.0  &  61.994966810(9)    &  -1.99448(5)    &   2.7(1.0)  \\
24\tablenotemark{a}  &  53978  &  53961.0  &  53995.7  &  61.993451919(5)    &  -1.99317(9)    &   \ldots   \\ 
25     &  54037  &   54002.4  &  54071.7  &  61.992457245(6)    &  -1.99447(3)    &   0.7(6)    \\
26     &  54175  &   54088.4  &  54263.0  &  61.9901021674(12)  &  -1.994236(3)   &  0.809(17)   \\ 
27     &  54357  &   54272.0  &  54442.3  &  61.986996430(3)    &  -1.994464(5)   &   0.80(4)  \\ 
28     &  54496  &   54455.0  &  54534.2  &  61.984616030(5)    &  -1.994663(18)  &   2.0(3)   \\ 
29\tablenotemark{a}  &  54557  &  54541.7  &  54573.4  &  61.983571888(5)    &  -1.99501(13)   &   \ldots   \\ 
30\tablenotemark{a}  &  54597  &  54579.1  &  54616.1  &  61.982891536(3)    &  -1.99508(7)    &   \ldots   \\ 
31     &  54675  &   54640.2  &  54710.2  &  61.981555135(4)    &  -1.99476(2)    &   2.7(4)    \\
32     &  54736  &   54714.2  &  54758.5  &  61.980510421(10)   &  -1.99502(8)    &   2(2)     \\ 
33     &  54828  &   54770.7  &  54885.4  &  61.978947039(3)    &  -1.994944(6)   &   0.91(8)  \\ 
34     &  54970  &   54896.7  &  55040.7  &  61.9765206710(17)  &  -1.994814(4)   &   0.95(3)   \\
35     &  55113  &   55048.8  &  55181.7  &  61.974069468(2)    &  -1.994486(6)   &   2.16(6)   \\
36\tablenotemark{a}  &  55196  &  55185.4  &  55208.0  &  61.972652440(6)    &  -1.9950(2)     &   \ldots   \\ 
37     &  55353  &   55275.3  &  55431.9  &  61.9699800967(11)  &  -1.995126(3)   &   0.62(2)   \\
38     &  55487  &   55457.8  &  55502.8  &  61.967680827(4)    &  -1.99511(5)    &  2.9(1.5)  \\ 
39     &  55530  &   55511.6  &  55549.3  &  61.966947205(9)    &  -1.99559(9)    &  3(3)      \\ 
40\tablenotemark{a}     &  55573  &  55562.5  &  55584.5  &  61.966206682(6)    &  -1.9942(2)     &  \ldots   \\ 
41\tablenotemark{a}  &  55599  &  55589.2  &  55610.6  &  61.965763933(8)    &  -1.9958(2)     &  \ldots   \\ 
42     &  55730  &   55627.3  &  55810.9  &  61.9635333927(16)  &  -1.995122(7)   &  0.73(2)   \\ 
43     &  55872  &  55818.5  &  55927.0  &  61.961105075(3)    &  -1.995471(7)   &   1.19(9) \\
\enddata
\tablenotetext{a}{These data segments are too short in duration to fit for a significant value of $\ddot{\nu}$.}
\tablenotetext{b}{The reference epoch for this segment of data (MJD $52717.3-52728.1$) is identical to the previous span; this is because its particularly short duration made it difficult to find a timing solution with a reference epoch centred within the 10.8-day time span of this segment.}
\end{deluxetable*}

%% file: 0537_glitch_table.tex
\startlongtable
\begin{deluxetable*}{llllll}
\tablecolumns{6}
\tablecaption{Observed glitches from PSR~J0537$-$6910.\label{tab:0537_glitches}}
\tablewidth{0pc}
\tablehead{
    \colhead{Glitch No.}  &  
    \colhead{Glitch epoch (MJD)}  &   
    \colhead{$\Delta\nu$\, $(\times 10^{-5}$\,s$^{-1})$}    &   
    \colhead{$\frac{\Delta\nu}{\nu}\, (\times 10^{-7})$}    &   
    \colhead{$\Delta\dot{\nu}$\, $(\times 10^{-13}$\,s$^{-2})$}  &
    \colhead{$\frac{\Delta\dot{\nu}}{\dot{\nu}}\, (\times 10^{-4})$}
}
\startdata 
1    &  51287(24)      &       4.2178(18)   &    6.799(3)    &      -1.16(11)    &    5.8(5)    \\ 
2    &  51562(15)      &       2.79249(18)  &    4.5015(3)   &      -1.500(6)    &    7.53(3)   \\
3    &  51711(5)       &       1.9549(2)    &    3.1515(4)   &      -1.331(11)   &    6.68(6)   \\
4    &  51827(27)      &       0.863(3)     &    1.391(4)    &      -0.95(11)    &    4.8(5)    \\
5    &  51881(6)       &       0.8768(8)    &    1.4136(13)  &      -0.70(10)    &    3.5(5)    \\
6    &  51960(5)       &       2.8190(4)    &    4.5447(6)   &      -1.55(2)     &    7.78(11)  \\
7    &  52176(11)      &       1.1439(11)   &    1.8443(18)  &      -1.76(11)    &    8.8(6)    \\
8    &  52241(12)      &       2.6395(12)   &    4.256(2)    &      -0.89(12)    &    4.5(6)    \\
9    &  52386(4)       &       1.0402(5)    &    1.6772(8)   &      -1.16(4)     &    5.81(19)  \\
10   &  52453(7)       &       1.3482(7)    &    2.1738(12)  &      -0.67(6)     &    3.4(3)    \\
11   &  52534(5)       &       2.6186(3)    &    4.2224(5)   &      -0.748(19)   &    3.75(10)  \\
12   &  52716(1)       &       0.02(5)      &    0.03(8)     &     -0.5(7)      &    3(3)      \\       
13   &  52737(9)       &       0.89(6)      &    1.44(9)     &      -0.8(7)      &    4(3)      \\ 
14   &  52807(15)      &       1.593(3)     &    2.568(4)    &      -1.2(2)      &    5.8(1.1)  \\
15   &  52886(3)       &       1.4536(4)    &    2.3441(7)   &      -1.13(3)     &    5.66(13)  \\
16   &  53014(6)       &       2.0978(3)    &    3.3830(4)   &      -1.466(12)   &    7.36(6)   \\
17   &  53134(13)      &       2.5250(4)    &    4.0721(6)   &      -1.151(16)   &    5.77(8)   \\
18   &  53288(3)       &       2.4510(3)    &    3.9529(4)   &      -1.301(6)    &    6.53(3)   \\
19   &  53445.1(1.7)   &       1.6071(3)    &    2.5921(5)   &      -1.574(8)    &    7.90(4)   \\
20   &  53550.4(1.5)   &       1.9926(3)    &    3.2138(5)   &      -1.429(9)    &    7.17(4)   \\
21   &  53695(8)       &       2.5346(4)    &    4.0883(6)   &      -1.625(11)   &    8.15(5)   \\
22   &  53861.0(1.4)   &       1.4548(9)    &    2.3466(15)  &      -1.82(6)     &    9.1(3)    \\
23   &  53940(21)      &       0.087(4)     &    0.140(6)    &       0.1(3)      &   -0.7(1.3)  \\
24   &  53999(3)       &       2.1827(11)   &    3.5209(17)  &      -1.54(10)    &    7.7(5)    \\
25   &  54080(8)       &       2.2998(10)   &    3.7099(16)  &      -0.70(7)     &    3.5(4)    \\
26   &  54268(5)       &       3.0315(3)    &    4.8905(5)   &      -1.492(7)    &    7.48(4)   \\
27   &  54449(6)       &       1.4871(7)    &    2.3991(11)  &      -1.66(3)     &    8.31(17)  \\
28   &  54538(4)       &       0.7050(11)   &    1.1374(18)  &      -1.08(13)    &    5.4(7)    \\
29   &  54576(3)       &       0.9136(10)   &    1.4739(16)  &      -0.07(15)    &    0.3(7)    \\
30   &  54628(12)      &       0.8211(13)   &    1.325(2)    &      -0.76(9)     &    3.8(5)    \\
31   &  54712(2)       &       0.6572(12)   &    1.060(2)    &      -1.63(12)    &    8.2(6)    \\
32   &  54765(6)       &       2.2438(17)   &    3.620(3)    &      -1.04(18)    &    5.2(9)    \\
33   &  54891(6)       &       2.1186(4)    &    3.4184(6)   &      -1.015(11)   &    5.09(6)   \\
34   &  55045(4)       &       1.3413(3)    &    2.1643(5)   &      -1.560(10)   &    7.82(5)   \\
35   &  55183.6(1.9)   &       1.2917(10)   &    2.0842(16)  &      -1.9(2)      &    9.4(1.1)  \\
36   &  55242(34)      &       3.426(7)     &    5.528(11)   &      -0.7(2)      &    3.4(1.1)  \\
37   &  55445(13)      &       1.059(2)     &    1.710(3)    &      -1.5(2)      &    7.6(1.1)  \\
38   &  55507(4)       &       0.7707(16)   &    1.244(3)    &      -1.6(2)      &    7.8(1.1)  \\
39   &  55556(7)       &       0.061(3)     &    0.098(4)    &      0.7(3)       &   -3.6(1.6)  \\ 
40   &  55587(2)       &       0.5405(17)   &    0.872(3)    &      -1.6(3)      &    8.1(1.6)  \\ 
41   &  55619(8)       &       2.808(3)     &    4.532(4)    &       0.02(24)    &   -0.1(1.2)  \\
42   &  55815(4)       &       1.9580(3)    &    3.1600(6)   &      -1.474(11)   &    7.39(6)  \\
\enddata
\end{deluxetable*}